\documentclass[twocolumn]{aastex63}
\usepackage{amsmath,amstext,amssymb}
\usepackage{xspace}
\usepackage{lipsum}
\usepackage{relsize}
\usepackage{appendix}

\newcommand{\Milwaukee}{University of Wisconsin-Milwaukee, Milwaukee, WI 53201, USA}
\newcommand{\CMU}{McWilliams Center for Cosmology, Department of Physics, Carnegie Mellon University, Pittsburgh, PA 15213, USA}

\shortauthors{Maga\~na~Hernandez and Ray}

\begin{document}

\title{Beyond Gaps and Bumps: Spectral Siren Cosmology with Non-Parametric Population Models}
\shorttitle{Beyond Gaps and Bumps}

\author[0000-0003-2362-0459]{Ignacio Maga\~na~Hernandez}
\affiliation{\CMU}
\email{imhernan@andrew.cmu.edu}

\author[0000-0002-7322-4748]{Anarya Ray}
\affiliation{\Milwaukee}
\email{anarya@uwm.edu}

\newcommand{\kmsMpc}{\ensuremath{\mbox{km s}^{-1} \,\mbox{Mpc}^{-1}}\xspace}
\newcommand{\Mpc}{\ensuremath{\mbox{Mpc}}\xspace}
\newcommand{\sqdeg}{\ensuremath{\text{deg}^2}\xspace}



\newcommand{\HubbleCombined}{\ensuremath{H_0=73.0^{+13.3}_{-7.7}}\xspace}
\newcommand{\HubbleBNS}{\ensuremath{H_0=71^{+23}_{-8}}\xspace}

\newcommand{\HubbleBBH}{\ensuremath{H_0=82.2^{+51.7}_{-33.7}}\xspace}
\newcommand{\matter}{\ensuremath{\Omega_m=0.3^{+0.1}_{-0.1}}\xspace}
\newcommand{\lengthscalem}{\ensuremath{l_m=2.1^{+1.0}_{-1.1}}\xspace}
\newcommand{\lengthscalez}{\ensuremath{l_z=2.1^{+1.0}_{-1.1}}\xspace}

\newcommand{\HubbleBBHSims}{\ensuremath{H_0=73.1^{+37.7}_{-19.9}}\xspace}
\newcommand{\matterSims}{\ensuremath{\Omega_m=0.3^{+0.2}_{-0.1}}\xspace}
\newcommand{\lengthscalemSims}{\ensuremath{l_m=0.9^{+0.6}_{-0.3}}\xspace}
\newcommand{\lengthscalezSims}{\ensuremath{l_z=0.6^{+0.5}_{-0.4}}\xspace}

\begin{abstract}
Gravitational wave standard sirens typically require electromagnetic (EM) data to obtain redshift information to constrain cosmology. Difficult to find EM counterparts for bright sirens and galaxy survey systematics for dark sirens make cosmological constraints with spectral sirens, a gravitational wave data-only approach, extremely appealing. In this work, we use the GWTC-3 BBH detections as spectral sirens to constrain the BBH population and the underlying cosmological expansion with a flexible model for the black hole mass spectrum. We use a binned Gaussian process to model the BBH mass distribution in the source frame without any astrophysical assumptions on the shape and or inclusion (or lack of) features that drive the cosmological constraints as the redshifted detector frame masses become consistent with the underlying astrophysical mass distribution features. For GWTC-3 we find a measurement on the Hubble constant of \HubbleCombined \kmsMpc at $68\%$ C.L. when combined with that obtained from the bright standard siren analysis with GW170817 and its associated host galaxy NGC 4993. We find an improved estimate for the Hubble constant of around a factor of 1.4 times better than the GW170817 measurement alone. We validate our nonparametric spectral siren approach with simulations and benchmark its scalability and constraining performance when compared with parametric methods. 
\end{abstract}


\section{Introduction} 
\label{section:introduction}

With the recent release of the third Gravitational-wave Transient Catalog \citep[GWTC-3]{LIGOScientific:2021djp} from the LIGO Scientific, Virgo, and KAGRA Collaborations \citep[LVK, ][]{LIGOScientific:2014pky,VIRGO:2014yos,KAGRA:2020agh,KAGRA:2013rdx}, the number of confident gravitational wave (GW) detections from compact binary mergers is around 70, most being from binary black hole (BBH) mergers. The increasing size of GW catalogs has enabled the study of the BBH population~\citep{LIGOScientific:2020kqk,LIGOScientific:2021psn}, the cosmic expansion history~\citep{LIGOScientific:2019zcs,LIGOScientific:2021aug} as well as allowing for extensive tests of general relativity (GR) within the strong field regime~\citep{KAGRA:2021duu,LIGOScientific:2021sio,LIGOScientific:2023bwz}. With half of the LVK's fourth observing run (O4) having concluded, the number of additional BBH merger detections has already doubled, and by its conclusion, we expect to have $\mathcal{O}(300)$ additional BBH mergers and $\mathcal{O}(10)$ mergers that contain at least one neutron star with $\mathcal{O}(1)$ being a multimessenger event. 

Gravitational waves have become a promising avenue to study the cosmic expansion history of the universe. Measurements of the Hubble constant ($H_0$) with local probes such as type 1A supernovae standard candles \citep{riess2022comprehensive} are in tension with early universe constraints from the cosmic microwave background (CMB) \citep{Aghanim:2018eyx}. These two state-of-the-art measurements provide an independent way of determining the Hubble constant at the percent level, however, there is currently a larger than 5-sigma tension between these constraints, implying either new physics or unaccounted sources of systematics \citep{riess2022comprehensive,DiValentino:2021izs}.

To potentially resolve this tension, one ideally needs a third independent cosmological probe. Gravitational wave sources are so-called standard sirens~\citep{Schutz:1986gp, Holz:2005df}, that is, they provide an absolute measurement for the luminosity distance to the source without the need for the cosmic distance ladder as a calibrating step. To do cosmology with GW sources, we require an independent measurement of the source redshift.

For \emph{bright sirens}, such as binary neutron star mergers we can attempt to detect the associated electromagnetic counterpart (EM) e.g., a kilonovae, and measure its redshift directly~\citep{LIGOScientific:2017vwq,LIGOScientific:2017adf}. For \emph{dark sirens}, such as binary black hole mergers or bright sirens without a detectable (or missed) EM counterpart, we can determine the redshift statistically using galaxy surveys as prior information on the potential host galaxies for these sources~\citep{Schutz:1986gp,DelPozzo:2011vcw,Nair:2018ign,Chen:2017rfc,LIGOScientific:2018gmd,Gray:2019ksv,DES:2019ccw,LIGOScientific:2019zcs,DES:2020nay,Mukherjee:2020hyn,Diaz:2021pem,Ghosh:2023ksl,Mukherjee:2020mha,Gray:2023wgj,Gray:2021sew}. The growing catalog of BBH mergers is thus critical for cosmological studies since multimessenger events have proven difficult to find.  

However, even without accessible electromagnetic information, either as EM counterparts for direct $H_0$ measurements or complete enough galaxy surveys, one can still make a statistical measurement of redshift using the features of the population distribution of compact binary mergers. Since we measure redshifted detector-frame masses, $m^{\rm det} = m(1+z)$, one can model the expected source frame mass distribution to estimate the redshift $z$. So called \emph{spectral sirens}~\citep{Ezquiaga:2021ayr,Ezquiaga:2022zkx} therefore allow for a direct measurement of the cosmic expansion history with gravitational wave data alone.

The mass spectrum of LIGO–Virgo–KAGRA events introduces at least five independent mass features~\citep{KAGRA:2021duu}: the upper and lower edges of the pair-instability supernova (PISN) gap~\citep{Woosley:2002zz,Heger:2001cd,Heger:2002by,Woosley:2016hmi,2019ApJ...878...49W}, the upper and lower edges of the neutron star–black hole gap~\citep{Ye:2022qoe}, and the minimum neutron star mass~\citep{Suwa:2018uni}. However, the location of these features is still uncertain, and some BBH formation channels might form events with masses in the gaps, e.g., hierarchical mergers~\citep{Gerosa:2021mno} leading to a more complex mass spectrum. By using the full mass distribution, degeneracies between mass evolution and cosmological evolution can be broken. This self-calibrating \emph{spectral siren} method has the potential to provide precision constraints of both cosmology and the potential evolution of the mass distribution. For recent works on cosmological inference that relies on the astrophysical mass distribution of CBCs see \cite{Mastrogiovanni:2021wsd,Mastrogiovanni:2023emh,Mastrogiovanni:2023zbw,Gray:2023wgj}

However, these studies~\citep{Farr:2019rap,Ezquiaga:2022zkx,Chen:2024gdn} have assumed \emph{a priori} some knowledge about the underlying astrophysical processes to forward model the CBC population with simple parametric population models. For example, modeling expected features in the mass spectrum which are then used to measure the cosmological expansion. However, \cite{Pierra:2023deu} has shown that incorrect assumptions regarding the shape of the mass-spectrum and its redshift evolution can lead to significant~($3\sigma$) biases in the inferred cosmological parameters. For more details regarding the redshift evolution of the mass spectrum and its impact on the associated cosmological inference, see \cite{Mukherjee:2021rtw,Karathanasis:2022rtr}.

Therefore, given the significant uncertainty regarding known BBH formation models and the new features found in the BBH mass spectrum~\citep{KAGRA:2021duu}, there is a need for a flexible and non-parametric approach to spectral siren cosmology. Examples of non-parametric population models include: autoregressive processes~\citep{autoreg1}, splines~\citep{spline0,spline1}, Gaussian mixture models~\citep{mixture1,structure1,structure2}, adaptive kernel density estimation ~\cite{adaptivekde}, maximum population likelihood~\citep{Paynenonparam}, Dirichlet processes~\citep{DelPozzononparam} and binned Gaussian processes (GPs,~\cite{Ray:2023upk, Mohite:2022pui, KAGRA:2021duu}).

In this work, we employ, for the first time, a flexible population model on the BBH population distribution and perform hierarchical Bayesian inference on the GWTC-3 BBH detections to measure the cosmic expansion history using the spectral siren methodology. We model the BBH population using the binned gaussian process (BGP) approach \citep{Ray:2023upk,Ray:2024hos,Mohite:2022pui,KAGRA:2021duu} to place constraints on the Hubble constant without assuming a particular shape on the mass distribution of BBH mergers. For a discussion regarding how our method compares with a parallel investigation on non-parametric spectral sirens~\citep{Farah:2024potat}, see Sec.~\ref{section:conclusion}.

This paper is organized as follows. In Section~\ref{section:methods}, we summarize the hierarchical Bayesian framework describing the BGP spectral siren model used in our analysis. In Section~\ref{section:results}, we present the main results of this paper by analyzing the GWTC-3 BBH observations with our framework. In Section~\ref{section:validation}, we test our methodology with a simulated population that mimicks the GWTC-3 population and current GW catalog sizes. Finally, in Section~\ref{section:conclusion} we provide a summary of our work and discuss future directions.

\section{Methods}
\label{section:methods}
In this section, we describe the hierarchical inference framework used to simultaneously infer the cosmological parameters with the shape of the CBC mass spectrum.

\subsection{Flexible population model}
\label{section:methods/pop}
Following \cite{Ray:2023upk, Mohite:2022pui}, we construct our flexible population model in the source frame as a piece-wise binned function in masses and redshift, such that,
\begin{align} \label{binned-pop-model}
    \frac{dN}{dm_1dm_2dz}(m_1,m_2,z|n^\gamma,n^{\alpha}) = \frac{n^{\gamma}_mn^{\alpha}_z}{m_1m_2}\frac{dV}{dz}T_{\rm{obs}}(1+z)^{-1},
\end{align}
where $n^{\gamma}_mn^{\alpha}_z$ is the merger rate density of CBCs per log component mass per co-moving volume per unit time in the $\gamma$th mass bin and the $\alpha$th redshift bin. While the model in Equation~\ref{binned-pop-model} is more flexible than the one used in \cite{Mohite:2022pui} which assumes no evolution of the merger rate, it is less so than the correlated model of \cite{Ray:2023upk} where the joint distribution of masses and redshifts were modelled using a single piece-wise binned function.

Keeping in mind that the primary objective of measuring the Hubble parameter independent of the uncertainties regarding CBC formation channels, and the large measurement uncertainties in the mass-redshift correlations reported by \cite{Ray:2023upk}, we expect the un-correlated mass-redshift model given by Equation ~\ref{binned-pop-model} to cause no significant biases in our results given the size of current datasets. However, as catalogs continue to grow, we hope to relax this restriction on the existence of mass-redshift correlations in future studies for which we will use the generalized correlated model of \cite{Ray:2023upk}.

To fit a population model such as Equation \ref{binned-pop-model} to GW observations of CBC systems, it is necessary to convert detector frame observables such as redshifted masses and luminosity distances into their corresponding source-frame counterparts using a fixed cosmological model. In the next subsection, following the spectral siren method of \cite{Farr:2019rap,Ezquiaga:2022zkx}, we describe how to vary and infer the cosmological parameters simultaneously with the population distribution in the context of our flexible model.

\subsection{Simultaneous cosmological inference}
\label{section:methods/popcosmo}
Given a cosmological model characterized by parameters $\Omega$, it is possible to express the population model described in the previous section as a function of detector frame CBC observables instead,
\begin{align} \label{binned-popcosmo-model}
    &\frac{dN}{dm_1dm_2dz}(m^d_1,m^d_2,D_L|n^\gamma_m,n^{\alpha}_z,\Omega) \nonumber \\&= \frac{n^{\gamma}_mn^{\alpha}_z}{m^d_1m^d_2}\frac{dV_c}{dz}\left(D_L|\Omega\right)T_{\rm{obs}}\left(1+z(D_L|\Omega)\right),
\end{align}
where $m_1^d$, $m_2^d$ and $D_L$ are the observed detector frame masses and luminosity distances of CBCs and $z(D_L|\Omega)$ is the redshift of the source as a function of luminosity distance given a set of cosmological parameters. As usual, $dV_c/dz$ is the differential uniform-in-comoving volume element, $T_\mathrm{obs}$ is the total observation time, and the extra factor of $1/(1+z)$ converts source-frame time to detector-frame time.

By modeling the occurrence of CBCs as realizations of an inhomogeneous Poisson process, it is possible to infer hyper-parameters characterizing the spectral siren model of Equation~\ref{binned-popcosmo-model} from the posterior samples of masses and luminosity distances of each observed CBC using Bayesian hierarchical inference~\citep{Mandel:2018mve, Loredo:2004nn,Vitale:2020aaz}. Within this framework, the likelihood of population and cosmological hyper-parameters given GW data~($d$) from all events in an observed catalog can be constructed as,
\begin{align}
    &p(\vec{d}|\vec{n}_m,\vec{n}_z,\Omega)\nonumber \\ &=e^{-N_{\rm{det}}(\vec{n}_m,\vec{n}_z,\Omega)}\prod_i^{N_{\rm{obs}}} \left\langle \frac{\frac{dN}{dm_1dm_2dz}(\vec{n}_m,\vec{n_z},\Omega)}{p_{\text{PE}}(m_1,m_2,z)} \right\rangle_{\text{samples},i}\label{likelihood},
\end{align}
where $\left\langle \cdot \right\rangle_{{\rm{samples}},i}$ represents a Monte Carlo sum over posterior samples from the ith event reweighted by the prior used in its parameter estimation and $N_{\rm{det}}$ is the expected number of detections as a function of hyper-parameters.

Since the criteria for detection takes the form of a threshold imposed on some statistic such as false alarm rate, the estimation of $N_{\rm{det}}$ has to account for Malmquist biases resulting from the imposition of the detection criteria~\citep{Mandel:2018mve, Loredo:2004nn}. This is implemented by simulating a fiducial population of CBCs and injecting them into detector noise realizations~\cite{Essick:2022ojx,2019RNAAS...3...66F}. The set of simulated sources that are detected above thresholds can be re-weighted to the population model of Equation~\ref{binned-popcosmo-model} to estimate $N_{\rm{det}}(\vec{n}_m,\vec{n}_z,\Omega)$ in the following way:
\begin{align}
    &N_{\rm{det}}(\vec{n}_m,\vec{n}_z,\Omega) \nonumber\\&= \frac{K_{\rm{det}}}{K_{\rm{draw}}}\left\langle \frac{\frac{dN}{dm_1dm_2dz}(\vec{n}_m,\vec{n}_z,\Omega)}{p_{\rm{draw}}(m_1,m_2,z)} \right\rangle_{\text{samples,det}}\label{det}
\end{align}
where similarly, $\left\langle \cdot \right\rangle_{\text{samples,det}}$ represents a Monte Carlo sum over the parameters of detectable events re-weighted by the population used to generate the mock simulations. For a discussion regarding the convergence of Monte Carlo sums used to evaluate various terms in the likelihood see Appendix~\ref{sec:app:neff}

The likelihood defined in Equation~\ref{likelihood} can then be used to inform the shape of the CBC mass spectrum, the redshift evolution of the merger rate and the cosmological parameters self-consistently given a catalog of GW events. 

\subsection{Gaussian process prior and HMC sampling}
To infer the shape of the mass distribution using the likelihood of Eq.~\eqref{likelihood}, we draw the rate densities in each mass bin from a Gaussian process prior~\citep{Ray:2023upk,Mohite:2022pui} so that the posterior on the rate densities is given by,
\begin{align}
&p(\vec{n}_m,\vec{n}_z,\vec{\mu},\vec{\sigma},\vec{l}|\vec{d},\Omega) \propto p(\vec{d}|\vec{n}_m,\vec{n}_z,\Omega) \\ & \times p(\vec{\mu},\vec{\sigma},\vec{l})p(\vec{n}_m|\vec{\mu}_m,\sigma_m,l_m)\nonumber \\ & \times p(\vec{n}_z|\vec{\mu}_z,\sigma_z,l_z)\nonumber \label{posterior}
\end{align}
where $p(\vec{n}|\vec{\mu},\sigma,\lambda)$ is the GP prior and $\vec{\mu}$ is the mean function of the GP while $l,\sigma$ are parameters that control the correlation length and amplitudes of the GP's covariance matrix.

The prior on the GP hyperparameters is defined by  $p(\vec{\mu},\sigma,l)$. Following previous implementations of the binned Gaussian process population analysis, we chose the priors on the GP hyper-parameters ($\vec{\mu},l,\sigma$) to be Normal, Log Normal, and Half Normal respectively while modeling the covariance matrix as an exponential quadratic function.

Along with the rate-densities, we simultaneously draw the cosmological parameters uniform priors and use Equation~\ref{likelihood} to sample the posterior distribution,
\begin{align}
    p(\vec{n}_m,\vec{n}_z,\Omega|\vec{d})\propto p(\vec{n}_m,\vec{n}_z,\vec{\mu},\vec{\sigma},\vec{l}|\vec{d},\Omega)p(\vec{\Omega})
\end{align}
The posterior samples of $\vec{\Omega}$ represent measurements of the cosmological expansion that have been marginalized over the uncertainties regarding the shape of the CBC mass spectrum without making any astrophysical assumptions about any features present in the mass spectrum. Simultaneously, the samples of $\vec{n}$ can then be used to reconstruct the shape of the CBC mass distribution in a data-driven way, while being free of any biases that might result from the uncertainties in the measurements of cosmological parameters.
\newpage
\section{Results for GWTC-3}
\label{section:results}
In this section, we present our BGP spectral siren constraints using GWTC-3 BBH mergers. Following the LVK GWTC-3 population analysis choices \cite{LIGOScientific:2021psn}, we make use only of the BBH events that pass a 1 per year IFAR threshold. That is, we analyze a total of 69 confidently detected BBH mergers and exclude GW190814 as a population outlier. For the events incorporated in our analysis, the parameter estimation samples of detector frame masses and luminosity distances are used to compute the posterior weights of Equation~\ref{likelihood}. We also use the publicly available LVK's GWTC-3 sensitivity estimate injection campaigns to compute Equation~\ref{det}~\citep{gwtc-3pe,gwtc-3inj}.

\begin{figure}[ht]
\begin{center}
\includegraphics[width=0.48\textwidth]{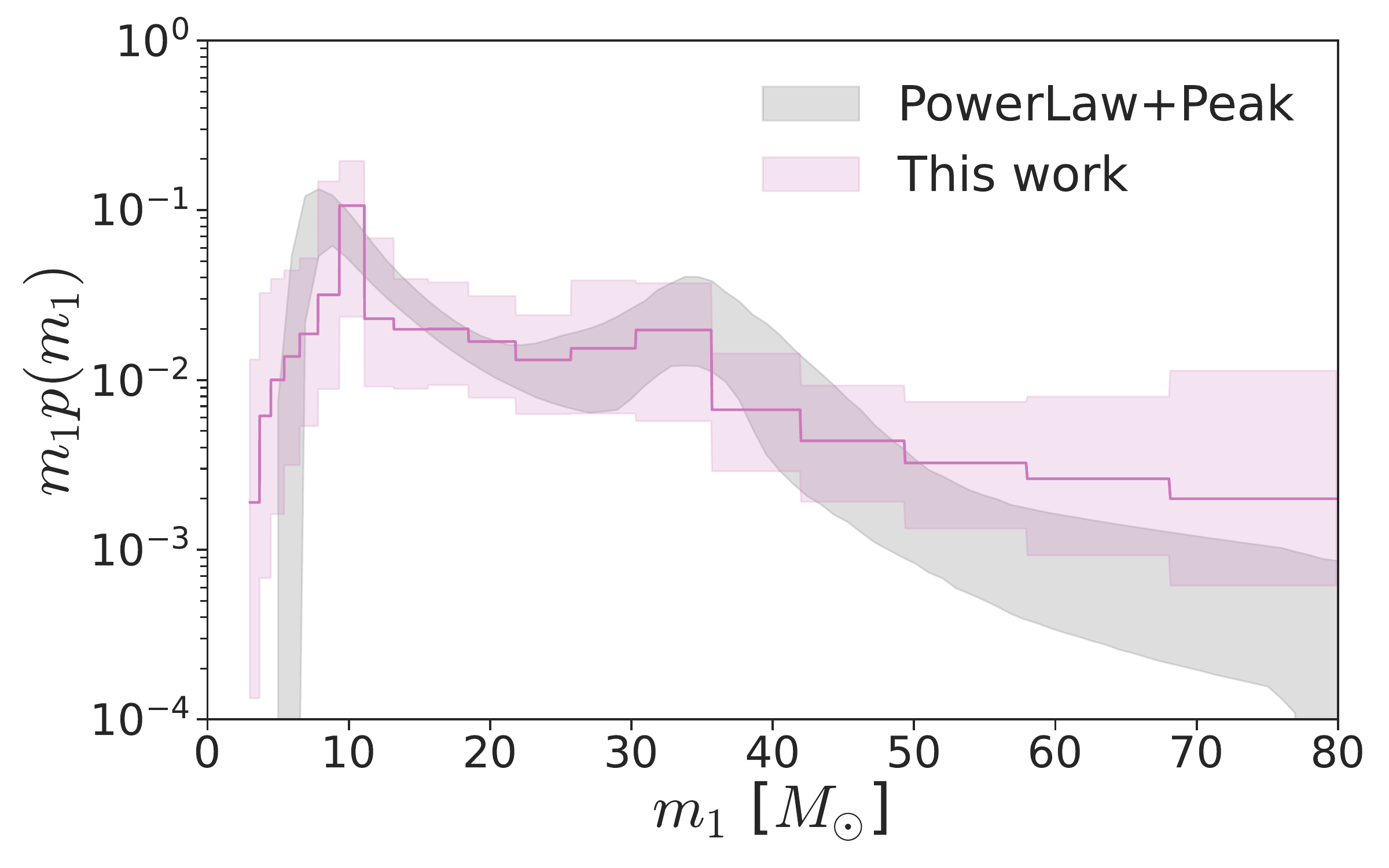}
\includegraphics[width=0.48\textwidth]{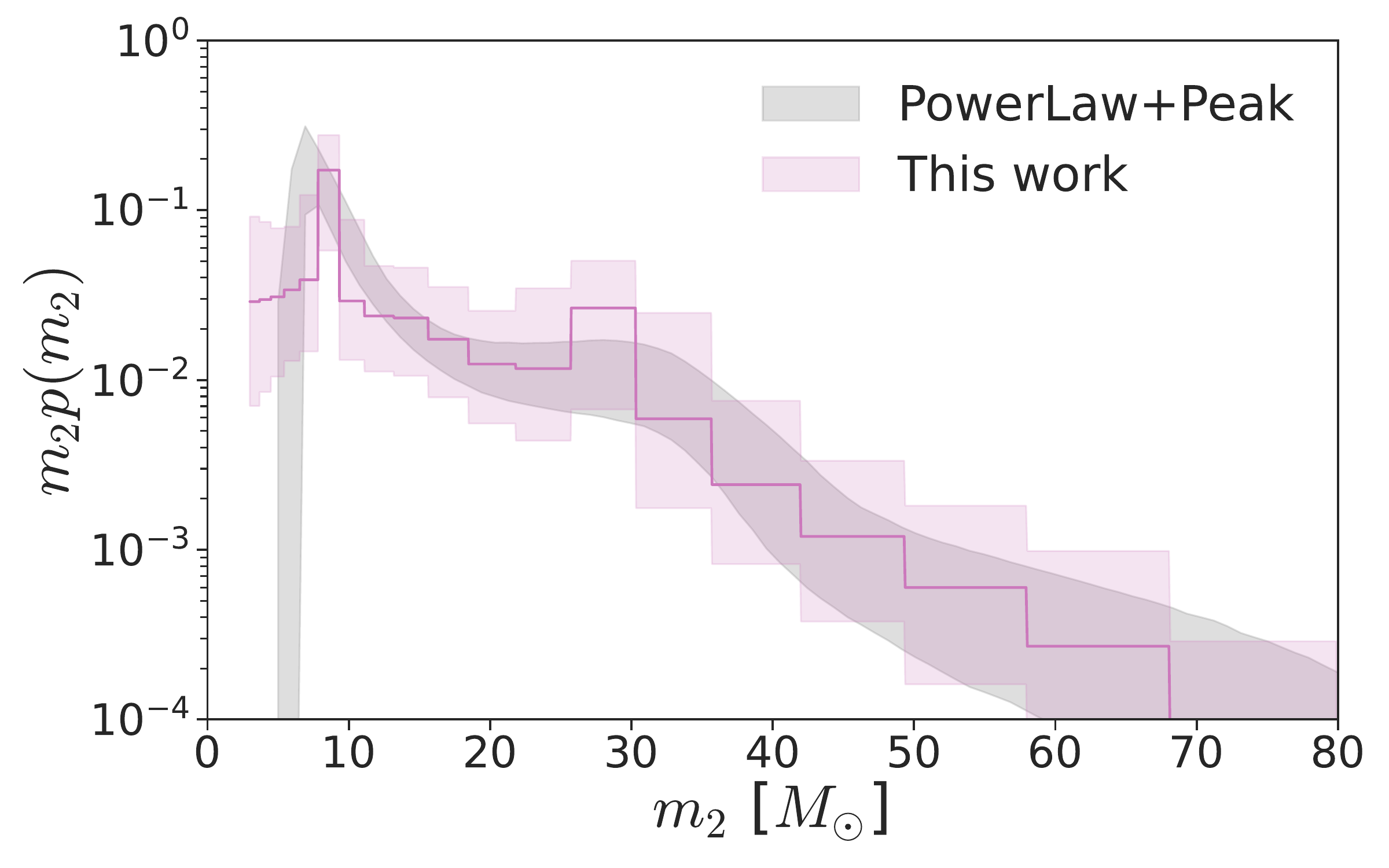}
\includegraphics[width=0.48\textwidth]{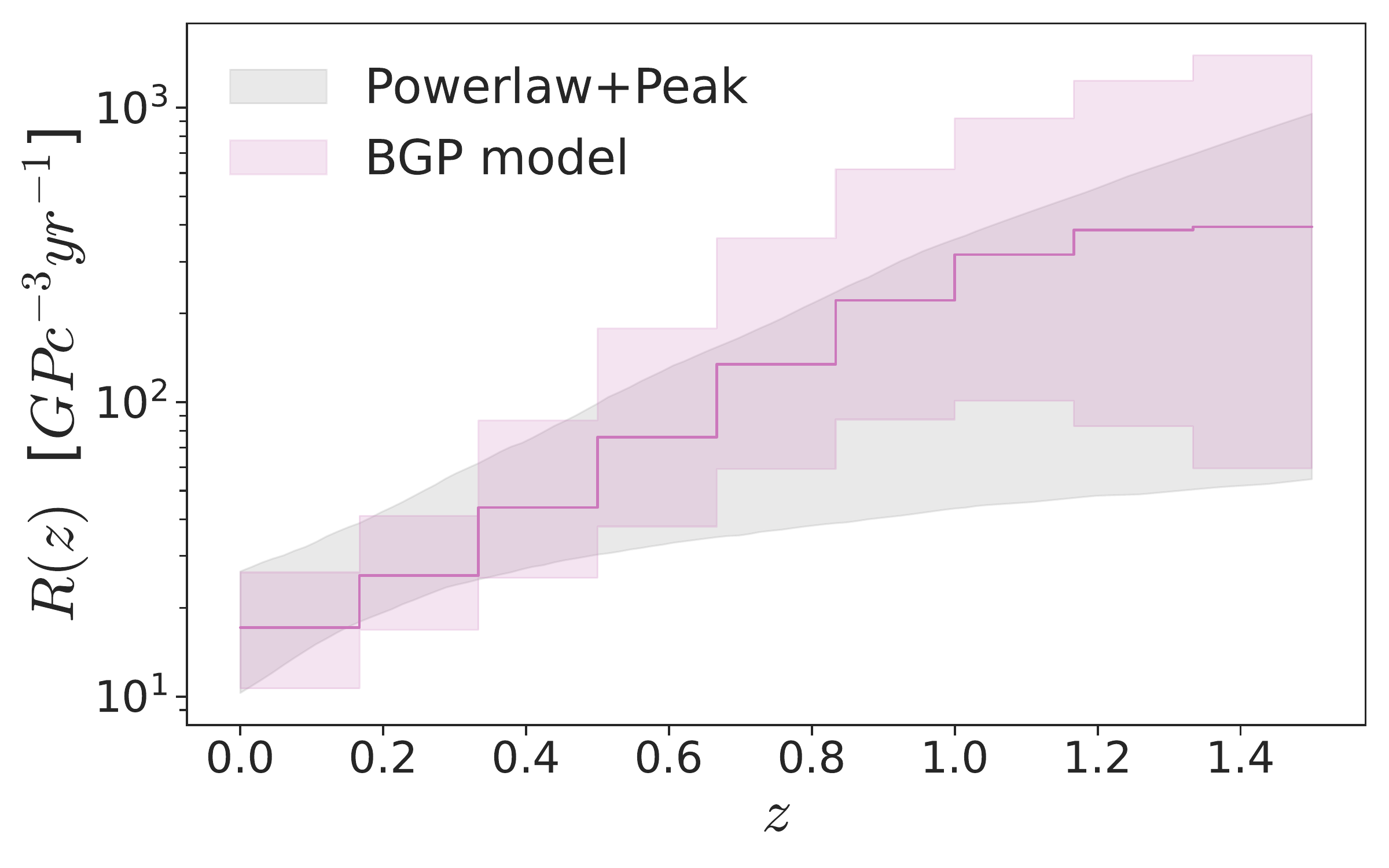}
\caption{\label{fig:mass} Inferred constraints for both the primary and secondary mass distributions using the 69 BBH mergers from GWTC-3 and our BGP spectral siren model. We show our results as the binned piecewise function plotted in purple and for comparison, we also show the corresponding results obtained by the LVK using the \textsc{Powerlaw + Peak} model in~\cite{LIGOScientific:2021psn} as the gray band depicting the 95\% posterior credible region.}
\end{center}
\end{figure}

We infer the mass rate densities as well as the redshift evolution of the BBH merger rate jointly with the cosmological parameters $H_0,\Omega_m$. In Figure~\ref{fig:mass}, we show the inferred marginal posterior distributions on both the primary mass $m_1$ and secondary mass $m_2$ along with the redshift evolution of the combined merger rate. We also show the corresponding LVK marginal posterior distributions on the masses for the \textsc{Powerlaw+Peak} model for reference~\cite{Talbot:2018cva,LIGOScientific:2021psn}. Our results show broad consistency with the LVK \textsc{Powerlaw+Peak} posterior. Our results are also consistent with the BGP results presented in \cite{Ray:2023upk} which assume a fixed cosmological model.

\begin{figure}[hb]
\begin{center}
\includegraphics[width=0.47\textwidth]{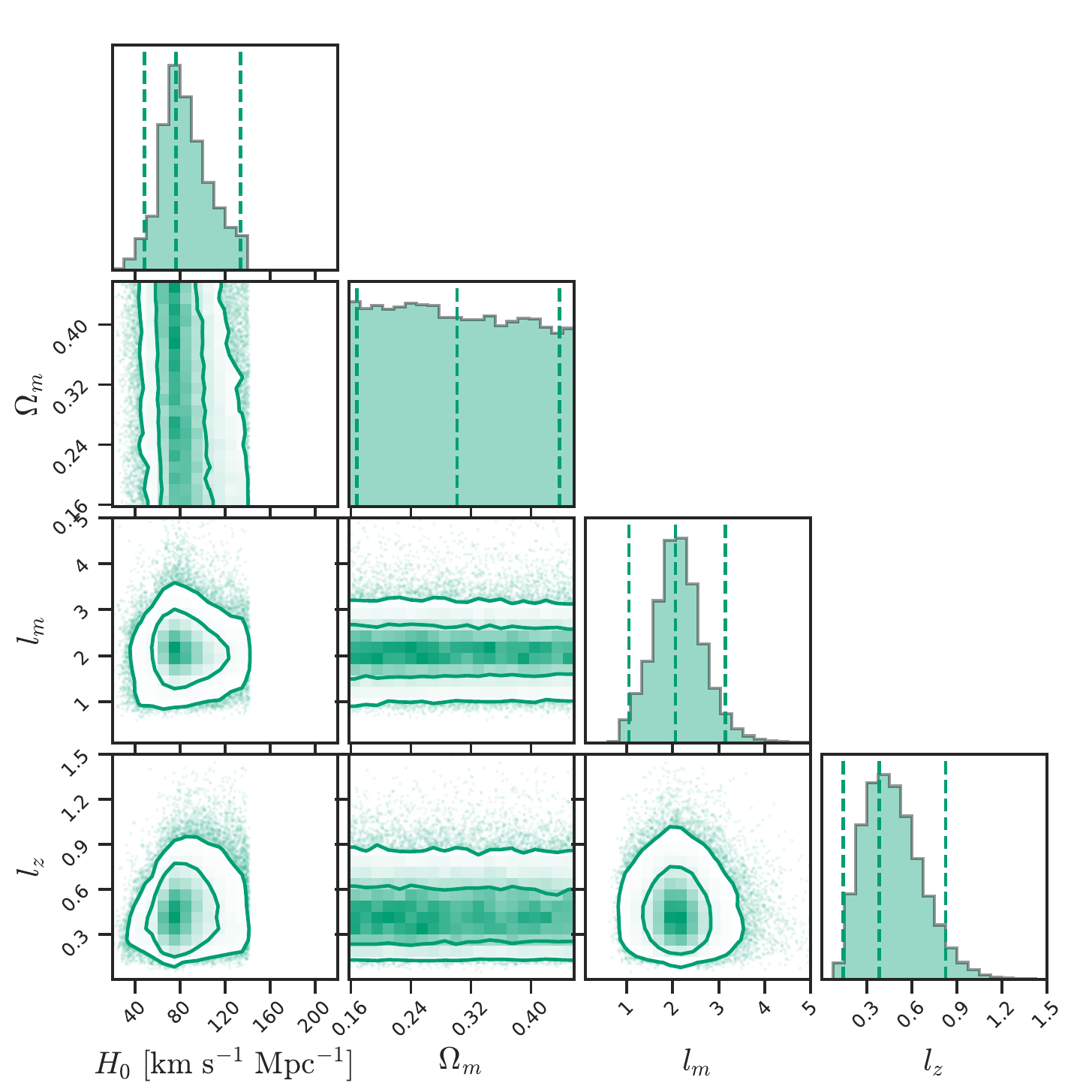}
\caption{\label{fig:corner} Corner plot showing the joint marginalized posterior distribution on the Hubble constant $H_0$, the matter density parameter $\Omega_m$, and the inferred length scales $l_m,l_z$ from our BGP spectral siren model and the 69 GWTC-3 BBH events.}
\end{center}
\end{figure}

In Figure \ref{fig:corner}, we show the corner plot for the joint posterior distribution on the Hubble constant $H_0$, the matter density parameter $\Omega_m$, as well as the GP kernel length scales $l_m,l_z$. We find a BGP spectral siren BBH measurement on the Hubble constant of \HubbleBBH \kmsMpc while for the matter density parameter, we find \matter both at $68\%$ C.L. We measure length scales \lengthscalem, \lengthscalez at $68\%$ C.L.

\begin{figure*}
\begin{center}
\includegraphics[width=0.9\textwidth]{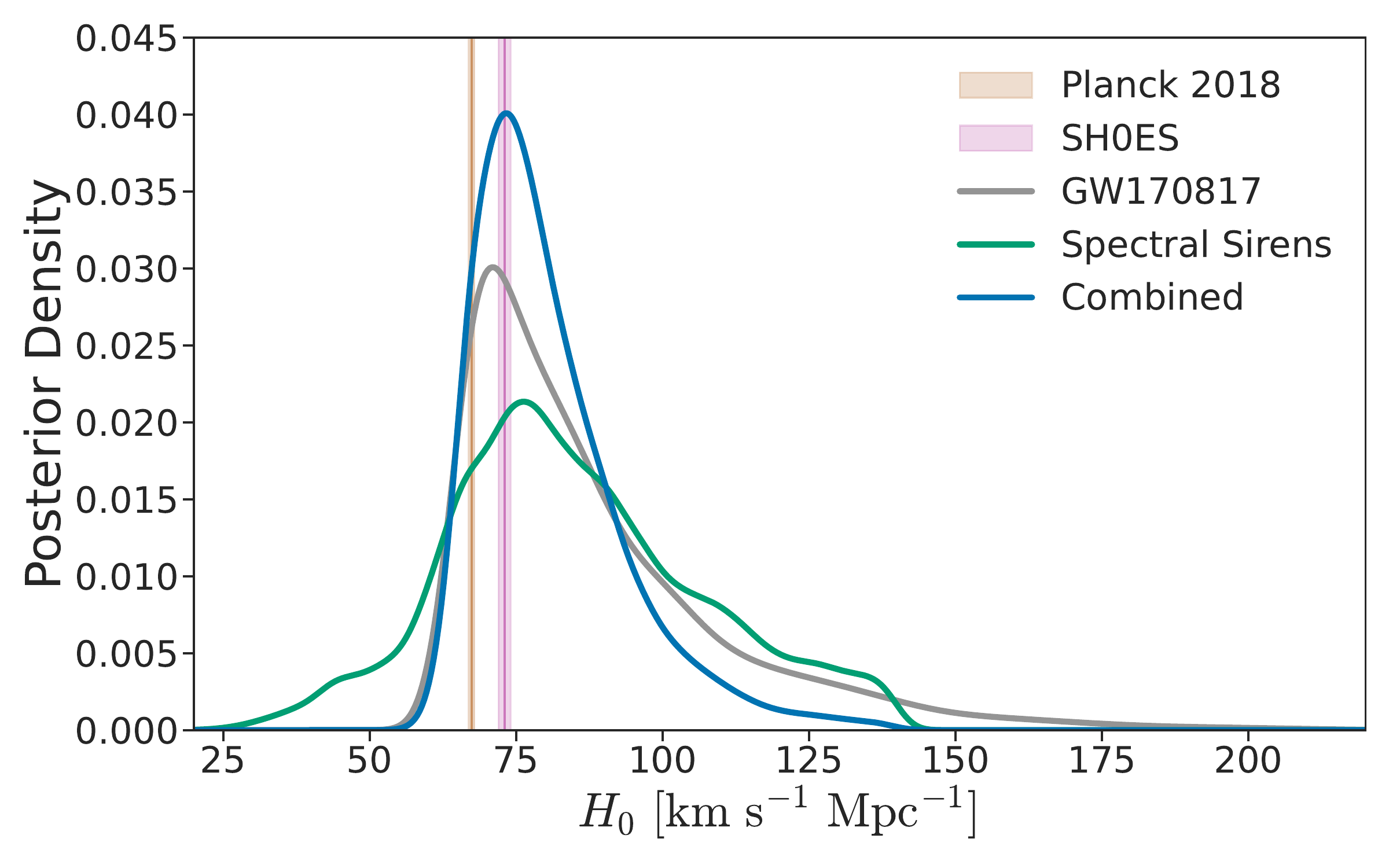}
\caption{\label{fig:hubble} Marginalized posterior distributions on the Hubble constant $H_0$. In green, we show the posterior from the 69 GWTC-3 BBH events we have considered in the spectral siren analysis using our BGP model. In gray, we show the GW170817 $H_0$ posterior obtained with its host galaxy NGC4993. In blue, we show the resulting combined GW170817 and GWTC-3 BBH spectral siren posterior to obtain a joint measurement on $H_0$. For reference, in the orange band we show measurements on $H_0$ from the CMB \citep{aghanim2020planck} and in the pink shaded band from standard candle type 1A SN measurements \citep{Riess:2019cxk}.}
\end{center}
\end{figure*}

In Figure \ref{fig:hubble}, we show joint constraints for our BGP spectral siren measurement on the Hubble constant with the 69 BBHs from GWTC-3 only when combined with the bright siren $H_0$ measurement from GW170817 and its EM counterpart NGC 4993~\citep{LIGOScientific:2017vwq,LIGOScientific:2017adf}. We find a joint $H_0$ measurement of \HubbleCombined \kmsMpc at $68\%$ C.L. which is an improvement with a factor of 1.5 times better relative to the GW180817 with NCG 4993 measurement alone (\HubbleBNS \kmsMpc at $68\%$ C.L.). In all of our reported $H_0$ measurements, we used a uniform prior in the range $[20,140] \ $ \kmsMpc.

\newpage
\section{Validation of Results}
\label{section:validation}
\begin{figure*}
\begin{center}
\includegraphics[width=0.32\textwidth]{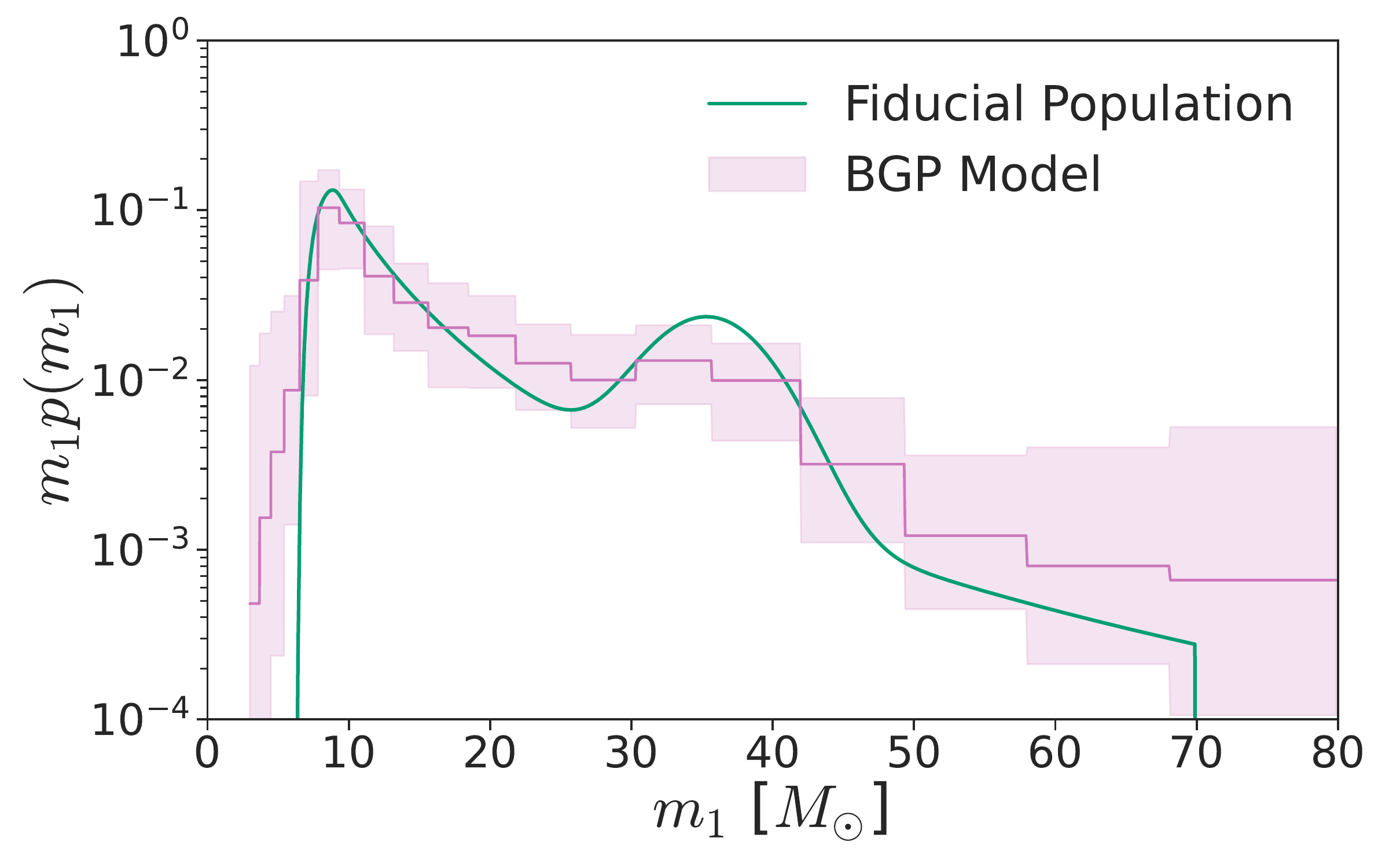}
\includegraphics[width=0.32\textwidth]{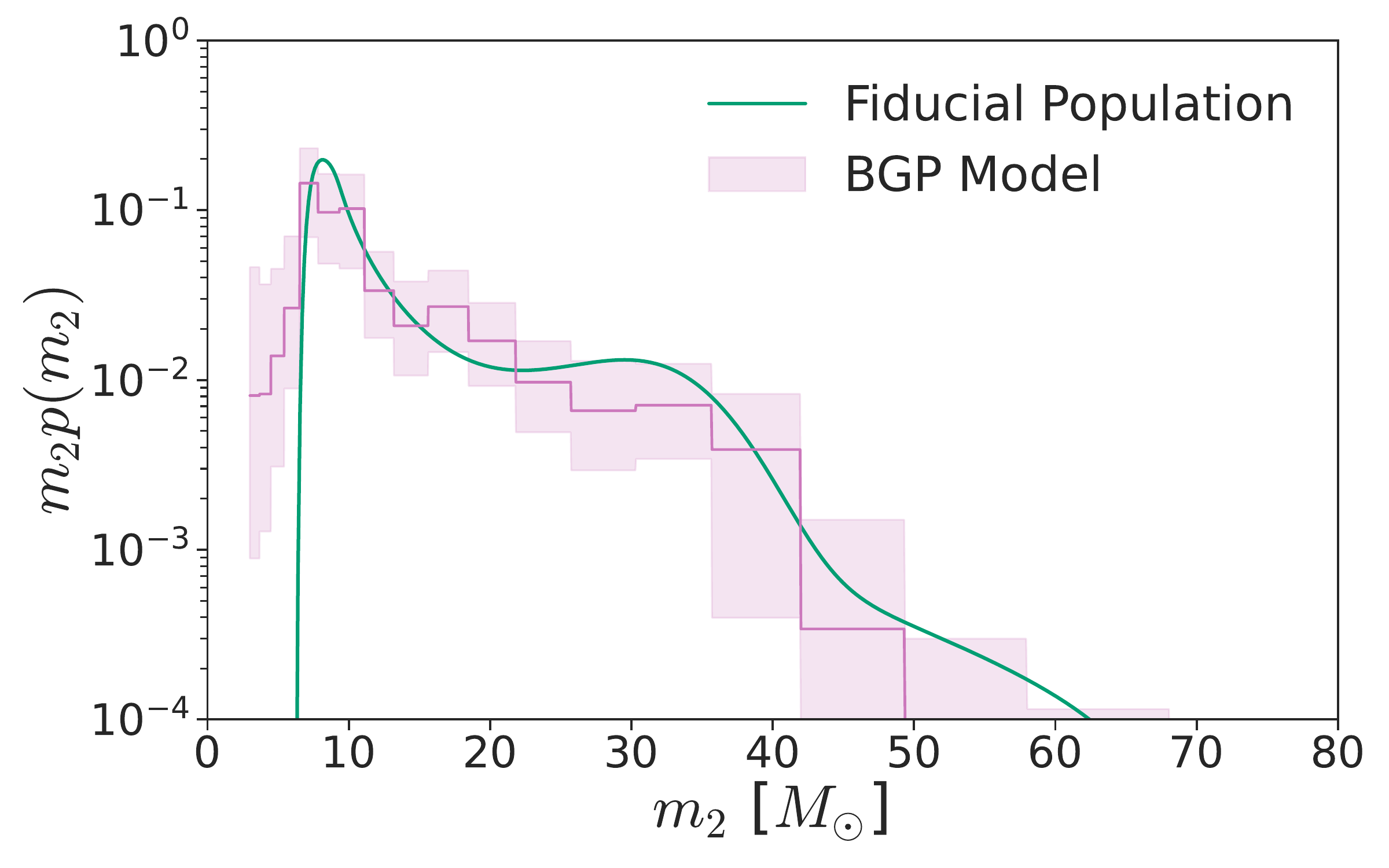}
\includegraphics[width=0.32\textwidth]{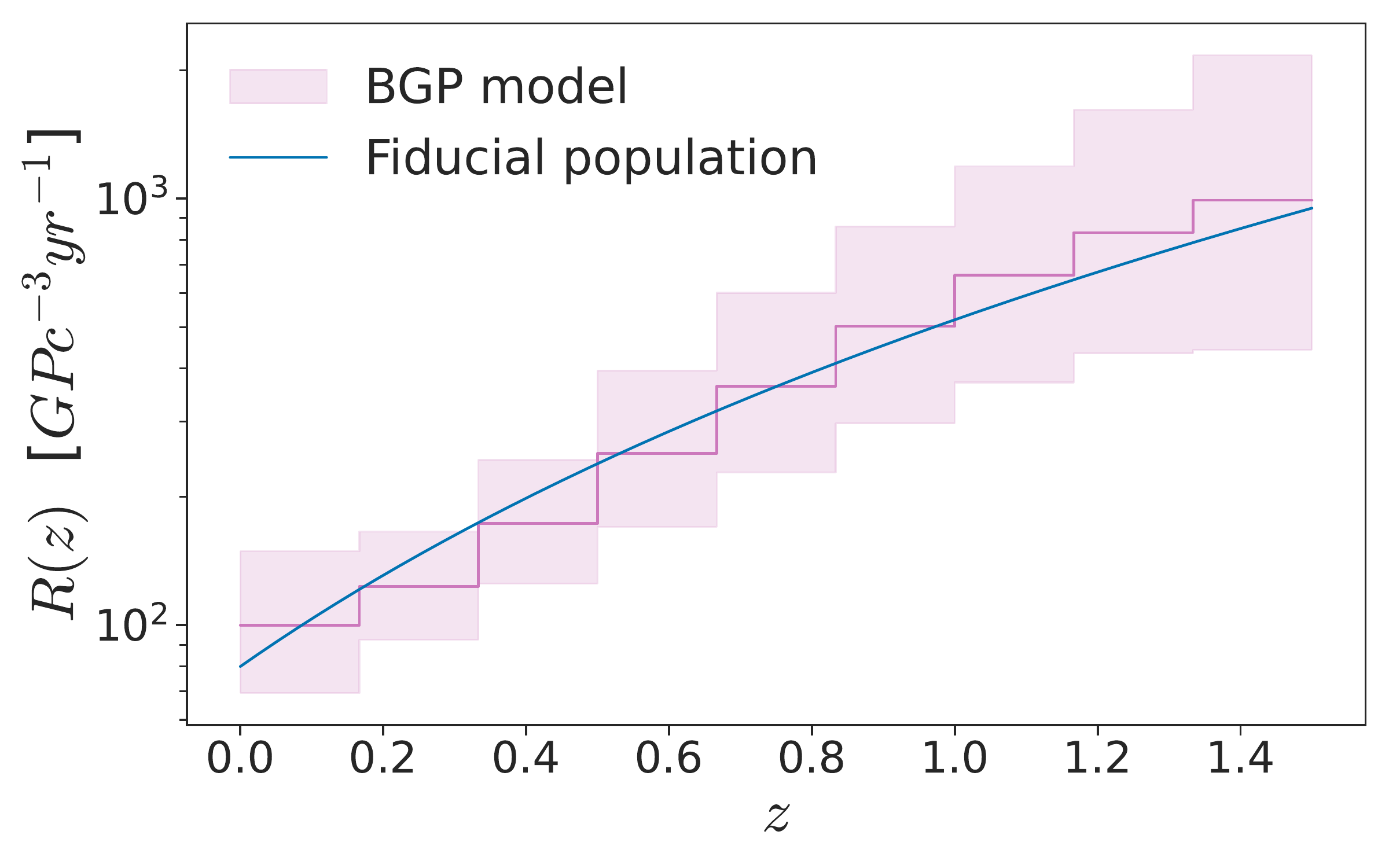}
\caption{\label{fig:mass_sims} Inferred constraints for both the primary and secondary mass distributions from the simulated population described in Section~\ref{section:validation} with the BGP spectral siren model. We show our results as the binned piecewise function plotted in purple. The simulated population is shown as the solid green lines. }
\end{center}
\end{figure*}

We validate our inference on simulated GW catalogs made from events drawn from a known fiducial population and cosmological model. We draw simulated BBH observations from the following models: for the primary mass, we draw from the \textsc{Powerlaw + Peak} model~\citep{Talbot:2018cva}, mass ratios are drawn from a powerlaw in $q$ with slope $\beta$~\citep{Fishbach:2019bbm} and we allow for the merger rate to evolve with redshift~\citep{redshift_ev}.
The fiducial hyperparameter values for the simulated population correspond to 
$m_{\rm{min}} = 5 M_{\odot}$, $m_{\rm{max}} = 65 M_{\odot}$, $\alpha=3.14$, $\beta=1.4$, $m_{{\rm peak}} = 35 M_{\odot}$, $\sigma_{{\rm peak}} = 5 M_{\odot}$, $f_{{\rm peak}} = 0.01$, and $\kappa=3$. For explicit expressions, we refer the reader to Appendix 1 of X and to Figure ~\ref{fig:mass_sims} for the marginal posterior distributions over $m_1$ and $m_2$ that we use in our simulations. We use a spatially flat $\Lambda$CDM cosmological model with assumed Planck 2015 cosmological parameters, that is, with $H_0 = 67.8 \ \rm{km \ s^{-1} \ {Mpc}^{-1}}$ and $\Omega_m = 0.308$ in our simulations.

The drawn masses and redshifts from the fiducial population are then assigned corresponding observed values calibrated to the expected measurement uncertainties for BBH mergers at advanced LIGO design sensitivity~\citep{Chen:2017rfc,KAGRA:2013rdx,Fishbach:2018edt} and selected so that only a subset is observable, i.e., simulating the selection effects seen in GW observations. The details of this procedure are described in~\cite{mockcat2,mockcat1,Fishbach:2018edt}. A corresponding set of simulated detectable injections is generated from a broad distribution to take into account the estimation of selection effects, a requirement for estimating $N_{\rm{det}}$ through Equation~\ref{det}.

Our catalog is comprised of 144 observed BBH events drawn from the fiducial population model and underlying cosmological model. As with our analysis of the GWTC-3 BBH events, we employ the BGP spectral siren model to fit our simulated population with the model defined in Equation~\ref{binned-popcosmo-model}. In Figure~\ref{fig:mass_sims} we show the inferred marginal posterior distributions on both the primary mass $m_1$ and secondary mass $m_2$ using our simulated data. We can see that the BGP model is sufficiently accurate at recovering the shape of the fiducial mass population with the 121 events that we consider. Although there is some discrepancy in the inferred shape of the primary mass population at high masses, we are still broadly consistent with the simulated population. We note that this discrepancy is due to a small number of events out of the total in our catalog being at the higher end of the mass spectrum as well as due to the increased uncertainty in their corresponding single-event parameter estimation posterior distributions.

\begin{figure}
\begin{center}
\includegraphics[width=0.47\textwidth]{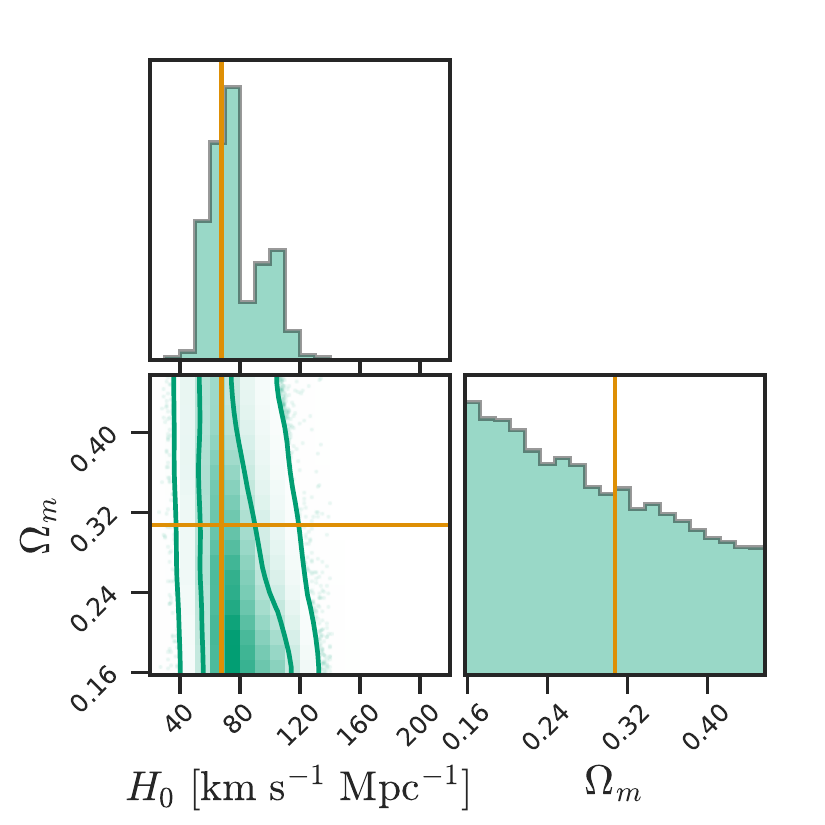}
\caption{\label{fig:corner_sims} Corner plot showing the joint posterior distributions on the Hubble constant $H_0$, the matter density parameter $\Omega_m$, and the inferred GP length scales from our BGP Spectral Siren model on the simulated population described in Section~\ref{section:validation}. We plot the simulated values for both $H_0 = 67.8 \ \rm{km \ s^{-1} \ {Mpc}^{-1}}$, $\Omega_m = 0.308$ for reference in gray.}
\end{center}
\end{figure}

We infer the BBH mass distribution as well as the redshift evolution of the BBH merger rate jointly with the Hubble constant $H_0$. In Figure \ref{fig:mass_sims}, we show the inferred marginal posterior distributions on both the primary mass $m_1$, secondary mass $m_2$, and redshift. We also show the fiducial populations for comparison. In figure~\ref{fig:corner_sims}, we show the posterior distributions of the cosmological parameters. The BGP spectral siren method can accurately reconstruct the shape of the injected population as well as the fiducial values of the cosmological parameters.


\section{Discussion and Conclusion}
In this work, we have explored how the spectral siren framework can be used jointly with nonparametric approaches in the modeling of the BBH population. In particular, we used a binned Gaussian process to model the black hole mass distribution without assuming any particular shape inclusive (or exclusive) of features expected from the astrophysics of BBH formation channels. Our proof of principle demonstration made use of the GWTC-3 confident BBH detections to place the first nonparametric constraints on the Hubble constant. We have validated our approach with GWTC-3 like populations and have made use of current catalog sizes to demonstrate the constraining power of our nonparametric method relative to traditional parametric approaches to spectral siren cosmology.

In a parallel investigation, \cite{Farah:2024potat} have also attempted to use Gaussian-process-based mass models on a simulated dataset for a proof of concept demonstration of astrophysics agnostic spectral siren cosmology. Even though their non-parametric population model is not restricted by bin resolutions, the mock data analysis presented in \cite{Farah:2024potat} models \emph{only} the primary mass distribution of CBCs with gaussian processes. They further assume that all BBHs are equal mass systems and use a parametric model for the redshift evolution of the merger rate. By simulating a large population of equal mass BBHs with a known functional form of the redshift evolution of the merger rate, they demonstrate that their method can implement accurate spectral siren cosmology in a simplistic Universe while being free of the necessity to bin up the space of BBH parameters. We note that while an unbinned population model in both component masses and redshifts could in principle be more flexible than ours, \cite{Farah:2024potat} have not demonstrated whether their method can be scalably generalized and applied to more realistic datasets wherein BBH masses are expected to follow more complicated pairing functions~\citep{KAGRA:2021duu} than a delta function in mass ratio centered around one. Furthermore, given the strong correlations between the measurement uncertainties of the Hubble parameter and the redshift evolution parameters~\citep{LIGOScientific:2021aug}, previously unmodelled features in the redshift evolution function can serve to equally bias the cosmological parameters. On the other hand, our model can reconstruct features in the joint distribution of component masses and the redshift evolution function completely driven by the data, up to the resolution limit imposed by our choice of binning, variations of which are found to yield consistent results.

As the size of GW catalogs grow, we expect to expand our nonparametric data-driven approach to take advantage of potential correlations between mass and redshift. In this work, we have assumed an uncorrelated mass-redshift distribution for simplicity. This choice can be relaxed as was shown in \cite{Ray:2023upk} where a single three dimensional BGP infers the shape of the joint mass-redshift distribution. However, as it was shown in \cite{Ray:2023upk}, most of the structure is washed out due to the increased dimensionality of the problem and thus would require larger GW catalogs to measure such correlations. This is important, if say for examples, the BBH mass distribution evolves at roughly the same rate as the cosmic expansion history as argued by \cite{Ezquiaga:2022zkx}. We leave these simulated and systematic studies as future work. 

Another caveat of our analysis is the fact that we only use BBH mergers to infer the cosmic expansion rate. Our methodology is CBC agnostic, that is, we fit for the merger rate density in mass bins. Including the whole population of CBC mergers (such as the inclusion of BNS and NSBH mergers) should make use of the full mass distribution of CBCs, a more powerful and event category agnostic spectral siren probe as explored by \cite{Ezquiaga:2022zkx}. 

Our method can also be expanded to be model agnostic concerning the underlying functional of the background cosmological model. In our work, we have assumed a spatially flat $\Lambda$CDM cosmological model which is sufficient given the number of observations that we have analyzed and their associated single event parameter uncertainties. Fitting the luminosity distance and redshift relation, $D_L(z)$, with a GP may provide a way to test the underlying cosmological model with future networks of GW detectors.

Finally, nonparametric methods are data-driven by nature and therefore require large amounts of observations for meaningful inference. As such, the scalability of such methods is critical. 
However, for the next generation of GW detectors (XG/3G), the development of scalable and potentially approximate inference will likely be necessary to make use of the  $\mathcal{O}(10^4 - 10^6)$ expected GW detections catalog sizes \citep{Chen:2024gdn}.

\label{section:conclusion}

\section*{Acknowledgements} 
The authors would like to thank Jolien Creighton for his useful comments and feedback throughout this work. IMH is supported by a McWilliams postdoctoral fellowship at Carnegie Mellon University. AR is supported by the National Science Foundation award PHY-2207728. The authors are grateful for computational resources provided by the LIGO Laboratory and supported by National Science Foundation Grants PHY-0757058 and PHY-0823459, PHY-0823459, PHY-1626190, and PHY-1700765. This material is based upon work supported by NSF's LIGO Laboratory which is a major facility fully funded by the National Science Foundation.
\appendix
\section{Convergence of Monte-Carlo integrals}
\label{sec:app:neff}
As mentioned in section~\ref{section:methods/popcosmo}, the Monte-Carlo sums used to compute various terms in Eq.~\eqref{likelihood} are susceptible to uncertainties which might raise concerns regarding their convergence. \cite{Farr:2019rap} shows that marginalizing over Monte Carlo uncertainties renders sampling of the marginalized posterior unable to find the maxima in the space of hyper-parameters unless sparsely populated regions of that space are penalized by means of a likelihood. Specifically, the effective number of single-event posterior samples and detectable injections are required to be \textit{high enough} compared to the total number of observations for unbiased inference. \cite{Essick:2022ojx,Farr:2019rap} outline how to impose this requirement in the context of an hierarchical framework such as the one employed in this study.

Following the implementations of the methods of \cite{Essick:2022ojx,Farr:2019rap} in the context of the binned model, as elaborated in \cite{Ray:2023upk}, we compute the means and standard deviations of the Monte Carlo sums used evaluate the likelihood of Eq.~\eqref{likelihood}. In particular, we compute the mean and deviations of the event specific posterior-weights and the detectable time-volumes in each bin as a function of the cosmological parameters, which take the following forms:
\begin{eqnarray}
    \sum_{\gamma,\alpha}n^{\gamma}_mn^{\alpha}_z\mu_{w,i}^{\gamma\alpha}(\Omega)&=&\left\langle \frac{\frac{dN}{dm_1dm_2dz}(\vec{n}_m,\vec{n}_z)}{p_{\text{PE}}(m_1,m_2,z)} \right\rangle_{\text{samples},i}\\
    \sum_{\gamma,\alpha}\left(n^{\gamma}_mn^{\alpha}_z\sigma_{w,i}^{\gamma\alpha}(\Omega)\right)^2&=&\left\langle\left( \frac{\frac{dN}{dm_1dm_2dz}(\vec{n}_m,\vec{n_z})}{p_{\text{PE}}(m_1,m_2,z)} \right)^2\right\rangle_{\text{samples},i}-\frac{1}{N_{\text{samples},i}}\left( \sum_{\gamma,\alpha}n^{\gamma}_mn^{\alpha}_z\mu_{w,i}^{\gamma\alpha}(\Omega)\right)^2\\
      \sum_{\gamma,\alpha}n^{\gamma}_mn^{\alpha}_z\mu_{\rm{VT}}^{\gamma\alpha}(\Omega)&=&\frac{K_{\text{det}}}{K_{\text{draw}}}\left\langle \frac{\frac{dN}{dm_1dm_2dz}(\vec{n}_m,\vec{n_z})}{p_{\text{draw}}(m_1,m_2,z)} \right\rangle_{\text{samples,det}}\\
    \sum_{\gamma,\alpha}\left(n^{\gamma}_mn^{\alpha}_z\sigma_{\rm{VT},i}^{\gamma}(\Omega)\right)^2&=&\left(\frac{K_{\text{det}}}{K_{\text{draw}}}\right)^2\left\langle\left( \frac{\frac{dN}{dm_1dm_2dz}(\vec{n}_m,\vec{n_z})}{p_{\text{draw}}(m_1,m_2,z)} \right)^2\right\rangle_{\text{samples, det}}-\left(\frac{K_{\text{det}}}{K_{\text{draw}}}\right)\left( \sum_{\gamma,\alpha}n^{\gamma}_mn^{\alpha}_z\mu_{\rm{VT}}^{\gamma\alpha}(\Omega)\right)^2
\end{eqnarray}
. The effective number of samples for the detectable injections and event-specific posteriors can be computed from these means and, deviations to be $N_{\text{eff}}=\left(\frac{\mu}{\sigma}\right)^2$ and are required to be large enough to avoid biases in the inferred hyper-parameters arising from Monte Carlo uncertainties. In particular, following \cite{Ray:2023upk,Callister:2023tgi}, we demand the following conditions:
\begin{eqnarray}
    \frac{N_{\rm{eff}}^{\rm{VT}}(\Omega)}{N_{\rm{det}}(\vec{n}_m,\vec{n_z},\Omega)}=\frac{\sum_{\gamma,\alpha}\left(\frac{\mu_{\rm{VT}}^{\gamma\alpha}}{\sigma_{\rm{VT}}^{\gamma\alpha}}\right)^2}{\sum_{\gamma,\alpha}n^{\gamma}_mn^{\alpha}_z\mu_{\rm{VT}}^{\gamma\alpha}}&\geq&2\\
  \min_i N_{\rm{eff}}^{w,i}(\vec{n}_m,\vec{n}_z,\Omega)= \min_i\left\{\frac{\left(\sum_{\gamma,\alpha}n^{\gamma}_mn^{\alpha}_z\mu_{w,i}^{\gamma\alpha}\right)^2}{\sum_{\gamma,\alpha}(n^{\gamma}_mn^{\alpha}_z\sigma_{w,i}^{\gamma\alpha})^2}\right\}&\geq& 10^{0.6}
\end{eqnarray}
be used on all of the hyperparameter samples that are accepted at each HMC step. Given the bin resolution and sample-sizes used in our studies we expect these conditions to be satisfied automatically. Hence, instead of imposing the above conditions in the form of a likelihood penalization, we verify post sampling that all of the hyper-parameter samples satisfy them.

\bibliography{references}{}

\begin{thebibliography}{}
\expandafter\ifx\csname natexlab\endcsname\relax\def\natexlab#1{#1}\fi
\providecommand{\url}[1]{\href{#1}{#1}}
\providecommand{\dodoi}[1]{doi:~\href{http://doi.org/#1}{\nolinkurl{#1}}}
\providecommand{\doeprint}[1]{\href{http://ascl.net/#1}{\nolinkurl{http://ascl.net/#1}}}
\providecommand{\doarXiv}[1]{\href{https://arxiv.org/abs/#1}{\nolinkurl{https://arxiv.org/abs/#1}}}

\bibitem[{Aasi {et~al.}(2015)}]{LIGOScientific:2014pky}
Aasi, J., {et~al.} 2015, Class. Quant. Grav., 32, 074001, \dodoi{10.1088/0264-9381/32/7/074001}

\bibitem[{Abbott {et~al.}(2017{\natexlab{a}})}]{LIGOScientific:2017vwq}
Abbott, B.~P., {et~al.} 2017{\natexlab{a}}, Phys. Rev. Lett., 119, 161101, \dodoi{10.1103/PhysRevLett.119.161101}

\bibitem[{Abbott {et~al.}(2017{\natexlab{b}})}]{LIGOScientific:2017adf}
---. 2017{\natexlab{b}}, Nature, 551, 85, \dodoi{10.1038/nature24471}

\bibitem[{Abbott {et~al.}(2018)}]{KAGRA:2013rdx}
---. 2018, Living Rev. Rel., 21, 3, \dodoi{10.1007/s41114-020-00026-9}

\bibitem[{Abbott {et~al.}(2021{\natexlab{a}})}]{LIGOScientific:2019zcs}
---. 2021{\natexlab{a}}, Astrophys. J., 909, 218, \dodoi{10.3847/1538-4357/abdcb7}

\bibitem[{Abbott {et~al.}(2021{\natexlab{b}})}]{LIGOScientific:2021djp}
Abbott, R., {et~al.} 2021{\natexlab{b}}.
\newblock \doarXiv{2111.03606}

\bibitem[{Abbott {et~al.}(2021{\natexlab{c}})}]{LIGOScientific:2020kqk}
---. 2021{\natexlab{c}}, Astrophys. J. Lett., 913, L7, \dodoi{10.3847/2041-8213/abe949}

\bibitem[{Abbott {et~al.}(2021{\natexlab{d}})}]{LIGOScientific:2021psn}
---. 2021{\natexlab{d}}.
\newblock \doarXiv{2111.03634}

\bibitem[{Abbott {et~al.}(2021{\natexlab{e}})}]{LIGOScientific:2021aug}
---. 2021{\natexlab{e}}.
\newblock \doarXiv{2111.03604}

\bibitem[{Abbott {et~al.}(2021{\natexlab{f}})}]{LIGOScientific:2021sio}
---. 2021{\natexlab{f}}.
\newblock \doarXiv{2112.06861}

\bibitem[{Abbott {et~al.}(2023{\natexlab{a}})}]{KAGRA:2021duu}
---. 2023{\natexlab{a}}, Phys. Rev. X, 13, 011048, \dodoi{10.1103/PhysRevX.13.011048}

\bibitem[{Abbott {et~al.}(2023{\natexlab{b}})}]{LIGOScientific:2023bwz}
---. 2023{\natexlab{b}}.
\newblock \doarXiv{2304.08393}

\bibitem[{Acernese {et~al.}(2015)}]{VIRGO:2014yos}
Acernese, F., {et~al.} 2015, Class. Quant. Grav., 32, 024001, \dodoi{10.1088/0264-9381/32/2/024001}

\bibitem[{Aghanim {et~al.}(2020)Aghanim, Akrami, Ashdown, Aumont, Baccigalupi, Ballardini, Banday, Barreiro, Bartolo, Basak, {et~al.}}]{aghanim2020planck}
Aghanim, N., Akrami, Y., Ashdown, M., {et~al.} 2020, Astronomy \& Astrophysics, 641, A6

\bibitem[{Akutsu {et~al.}(2021)}]{KAGRA:2020agh}
Akutsu, T., {et~al.} 2021, PTEP, 2021, 05A102, \dodoi{10.1093/ptep/ptab018}

\bibitem[{Callister \& Farr(2023{\natexlab{a}})}]{autoreg1}
Callister, T.~A., \& Farr, W.~M. 2023{\natexlab{a}}, A Parameter-Free Tour of the Binary Black Hole Population

\bibitem[{Callister \& Farr(2023{\natexlab{b}})}]{Callister:2023tgi}
---. 2023{\natexlab{b}}.
\newblock \doarXiv{2302.07289}

\bibitem[{Chen {et~al.}(2024)Chen, Ezquiaga, \& Gupta}]{Chen:2024gdn}
Chen, H.-Y., Ezquiaga, J.~M., \& Gupta, I. 2024.
\newblock \doarXiv{2402.03120}

\bibitem[{Chen {et~al.}(2018)Chen, Fishbach, \& Holz}]{Chen:2017rfc}
Chen, H.-Y., Fishbach, M., \& Holz, D.~E. 2018, Nature, 562, 545, \dodoi{10.1038/s41586-018-0606-0}

\bibitem[{Del~Pozzo(2012)}]{DelPozzo:2011vcw}
Del~Pozzo, W. 2012, Phys. Rev. D, 86, 043011, \dodoi{10.1103/PhysRevD.86.043011}

\bibitem[{Di~Valentino {et~al.}(2021)Di~Valentino, Mena, Pan, Visinelli, Yang, Melchiorri, Mota, Riess, \& Silk}]{DiValentino:2021izs}
Di~Valentino, E., Mena, O., Pan, S., {et~al.} 2021, Class. Quant. Grav., 38, 153001, \dodoi{10.1088/1361-6382/ac086d}

\bibitem[{Diaz \& Mukherjee(2022)}]{Diaz:2021pem}
Diaz, C.~C., \& Mukherjee, S. 2022, Mon. Not. Roy. Astron. Soc., 511, 2782, \dodoi{10.1093/mnras/stac208}

\bibitem[{Edelman {et~al.}(2022{\natexlab{a}})Edelman, Doctor, Godfrey, \& Farr}]{spline0}
Edelman, B., Doctor, Z., Godfrey, J., \& Farr, B. 2022{\natexlab{a}}, The Astrophysical Journal, 924, 101, \dodoi{10.3847/1538-4357/ac3667}

\bibitem[{Edelman {et~al.}(2022{\natexlab{b}})Edelman, Farr, \& Doctor}]{spline1}
Edelman, B., Farr, B., \& Doctor, Z. 2022{\natexlab{b}}, Cover Your Basis: Comprehensive Data-Driven Characterization of the Binary Black Hole Population

\bibitem[{Essick \& Farr(2022)}]{Essick:2022ojx}
Essick, R., \& Farr, W. 2022.
\newblock \doarXiv{2204.00461}

\bibitem[{Ezquiaga(2021)}]{Ezquiaga:2021ayr}
Ezquiaga, J.~M. 2021, Phys. Lett. B, 822, 136665, \dodoi{10.1016/j.physletb.2021.136665}

\bibitem[{Ezquiaga \& Holz(2022)}]{Ezquiaga:2022zkx}
Ezquiaga, J.~M., \& Holz, D.~E. 2022, Phys. Rev. Lett., 129, 061102, \dodoi{10.1103/PhysRevLett.129.061102}

\bibitem[{Farah {et~al.}(2024)Farah, Callister, Ezquiaga, Zevin, \& Holz}]{Farah:2024potat}
Farah, A.~M., Callister, T.~A., Ezquiaga, J.~M., Zevin, M., \& Holz, D.~E. 2024.
\newblock \doarXiv{2404.02210}

\bibitem[{Farah {et~al.}(2023)Farah, Edelman, Zevin, Fishbach, Ezquiaga, Farr, \& Holz}]{mockcat2}
Farah, A.~M., Edelman, B., Zevin, M., {et~al.} 2023, Things that might go bump in the night: Assessing structure in the binary black hole mass spectrum

\bibitem[{Farr(2019)}]{Farr:2019rap}
Farr, W.~M. 2019, Research Notes of the AAS, 3, 66, \dodoi{10.3847/2515-5172/ab1d5f}

\bibitem[{{Farr}(2019)}]{2019RNAAS...3...66F}
{Farr}, W.~M. 2019, Research Notes of the American Astronomical Society, 3, 66, \dodoi{10.3847/2515-5172/ab1d5f}

\bibitem[{Fishbach {et~al.}(2020)Fishbach, Farr, \& Holz}]{mockcat1}
Fishbach, M., Farr, W.~M., \& Holz, D.~E. 2020, The Astrophysical Journal Letters, 891, L31, \dodoi{10.3847/2041-8213/ab77c9}

\bibitem[{Fishbach \& Holz(2020)}]{Fishbach:2019bbm}
Fishbach, M., \& Holz, D.~E. 2020, Astrophys. J. Lett., 891, L27, \dodoi{10.3847/2041-8213/ab7247}

\bibitem[{Fishbach {et~al.}(2018{\natexlab{a}})Fishbach, Holz, \& Farr}]{redshift_ev}
Fishbach, M., Holz, D.~E., \& Farr, W.~M. 2018{\natexlab{a}}, The Astrophysical Journal Letters, 863, L41, \dodoi{10.3847/2041-8213/aad800}

\bibitem[{Fishbach {et~al.}(2018{\natexlab{b}})Fishbach, Holz, \& Farr}]{Fishbach:2018edt}
---. 2018{\natexlab{b}}, Astrophys. J., 863, L41, \dodoi{10.3847/2041-8213/aad800}

\bibitem[{Fishbach {et~al.}(2019)}]{LIGOScientific:2018gmd}
Fishbach, M., {et~al.} 2019, Astrophys. J. Lett., 871, L13, \dodoi{10.3847/2041-8213/aaf96e}

\bibitem[{Gerosa \& Fishbach(2021)}]{Gerosa:2021mno}
Gerosa, D., \& Fishbach, M. 2021, Nature Astron., 5, 749, \dodoi{10.1038/s41550-021-01398-w}

\bibitem[{Ghosh {et~al.}(2023)Ghosh, More, Bera, \& Bose}]{Ghosh:2023ksl}
Ghosh, T., More, S., Bera, S., \& Bose, S. 2023.
\newblock \doarXiv{2312.16305}

\bibitem[{Gray {et~al.}(2019)Gray, Magaña~Hernandez, Qi, Sur, {et~al.}}]{Gray:2019ksv}
Gray, R., Magaña~Hernandez, I., Qi, H., Sur, A., {et~al.} 2019.
\newblock \doarXiv{1908.06050}

\bibitem[{Gray {et~al.}(2022)Gray, Messenger, \& Veitch}]{Gray:2021sew}
Gray, R., Messenger, C., \& Veitch, J. 2022, Mon. Not. Roy. Astron. Soc., 512, 1127, \dodoi{10.1093/mnras/stac366}

\bibitem[{Gray {et~al.}(2023)}]{Gray:2023wgj}
Gray, R., {et~al.} 2023, JCAP, 12, 023, \dodoi{10.1088/1475-7516/2023/12/023}

\bibitem[{Heger {et~al.}(2003)Heger, Fryer, Woosley, Langer, \& Hartmann}]{Heger:2002by}
Heger, A., Fryer, C.~L., Woosley, S.~E., Langer, N., \& Hartmann, D.~H. 2003, Astrophys. J., 591, 288, \dodoi{10.1086/375341}

\bibitem[{Heger \& Woosley(2002)}]{Heger:2001cd}
Heger, A., \& Woosley, S.~E. 2002, Astrophys. J., 567, 532, \dodoi{10.1086/338487}

\bibitem[{Holz \& Hughes(2005)}]{Holz:2005df}
Holz, D.~E., \& Hughes, S.~A. 2005, Astrophys. J., 629, 15, \dodoi{10.1086/431341}

\bibitem[{Karathanasis {et~al.}(2023)Karathanasis, Mukherjee, \& Mastrogiovanni}]{Karathanasis:2022rtr}
Karathanasis, C., Mukherjee, S., \& Mastrogiovanni, S. 2023, Mon. Not. Roy. Astron. Soc., 523, 4539, \dodoi{10.1093/mnras/stad1373}

\bibitem[{Loredo(2004)}]{Loredo:2004nn}
Loredo, T.~J. 2004, AIP Conf. Proc., 735, 195, \dodoi{10.1063/1.1835214}

\bibitem[{Mandel {et~al.}(2019)Mandel, Farr, \& Gair}]{Mandel:2018mve}
Mandel, I., Farr, W.~M., \& Gair, J.~R. 2019, Mon. Not. Roy. Astron. Soc., 486, 1086, \dodoi{10.1093/mnras/stz896}

\bibitem[{Mastrogiovanni {et~al.}(2021)Mastrogiovanni, Leyde, Karathanasis, Chassande-Mottin, Steer, Gair, Ghosh, Gray, Mukherjee, \& Rinaldi}]{Mastrogiovanni:2021wsd}
Mastrogiovanni, S., Leyde, K., Karathanasis, C., {et~al.} 2021, Phys. Rev. D, 104, 062009, \dodoi{10.1103/PhysRevD.104.062009}

\bibitem[{Mastrogiovanni {et~al.}(2023)Mastrogiovanni, Laghi, Gray, Santoro, Ghosh, Karathanasis, Leyde, Steer, Perries, \& Pierra}]{Mastrogiovanni:2023emh}
Mastrogiovanni, S., Laghi, D., Gray, R., {et~al.} 2023, Phys. Rev. D, 108, 042002, \dodoi{10.1103/PhysRevD.108.042002}

\bibitem[{Mastrogiovanni {et~al.}(2024)Mastrogiovanni, Pierra, Perri\`es, Laghi, Caneva~Santoro, Ghosh, Gray, Karathanasis, \& Leyde}]{Mastrogiovanni:2023zbw}
Mastrogiovanni, S., Pierra, G., Perri\`es, S., {et~al.} 2024, Astron. Astrophys., 682, A167, \dodoi{10.1051/0004-6361/202347007}

\bibitem[{Mohite(2022)}]{Mohite:2022pui}
Mohite, S. 2022, PhD thesis, Wisconsin U., Milwaukee

\bibitem[{Mukherjee(2022)}]{Mukherjee:2021rtw}
Mukherjee, S. 2022, Mon. Not. Roy. Astron. Soc., 515, 5495, \dodoi{10.1093/mnras/stac2152}

\bibitem[{Mukherjee {et~al.}(2021{\natexlab{a}})Mukherjee, Wandelt, Nissanke, \& Silvestri}]{Mukherjee:2020hyn}
Mukherjee, S., Wandelt, B.~D., Nissanke, S.~M., \& Silvestri, A. 2021{\natexlab{a}}, Phys. Rev. D, 103, 043520, \dodoi{10.1103/PhysRevD.103.043520}

\bibitem[{Mukherjee {et~al.}(2021{\natexlab{b}})Mukherjee, Wandelt, \& Silk}]{Mukherjee:2020mha}
Mukherjee, S., Wandelt, B.~D., \& Silk, J. 2021{\natexlab{b}}, Mon. Not. Roy. Astron. Soc., 502, 1136, \dodoi{10.1093/mnras/stab001}

\bibitem[{Nair {et~al.}(2018)Nair, Bose, \& Saini}]{Nair:2018ign}
Nair, R., Bose, S., \& Saini, T.~D. 2018, Phys. Rev., D98, 023502, \dodoi{10.1103/PhysRevD.98.023502}

\bibitem[{Palmese {et~al.}(2020)}]{DES:2020nay}
Palmese, A., {et~al.} 2020, Astrophys. J. Lett., 900, L33, \dodoi{10.3847/2041-8213/abaeff}

\bibitem[{Payne \& Thrane(2023)}]{Paynenonparam}
Payne, E., \& Thrane, E. 2023, Phys. Rev. Res., 5, 023013, \dodoi{10.1103/PhysRevResearch.5.023013}

\bibitem[{Pierra {et~al.}(2023)Pierra, Mastrogiovanni, Perri\`es, \& Mapelli}]{Pierra:2023deu}
Pierra, G., Mastrogiovanni, S., Perri\`es, S., \& Mapelli, M. 2023.
\newblock \doarXiv{2312.11627}

\bibitem[{{Planck Collaboration}(2018)}]{Aghanim:2018eyx}
{Planck Collaboration}. 2018.
\newblock \doarXiv{1807.06209}

\bibitem[{Ray {et~al.}(2024)Ray, Maga\~na Hernandez, Breivik, \& Creighton}]{Ray:2024hos}
Ray, A., Maga\~na Hernandez, I., Breivik, K., \& Creighton, J. 2024.
\newblock \doarXiv{2404.03166}

\bibitem[{Ray {et~al.}(2023)Ray, Maga\~na Hernandez, Mohite, Creighton, \& Kapadia}]{Ray:2023upk}
Ray, A., Maga\~na Hernandez, I., Mohite, S., Creighton, J., \& Kapadia, S. 2023, Astrophys. J., 957, 37, \dodoi{10.3847/1538-4357/acf452}

\bibitem[{Riess {et~al.}(2019)Riess, Casertano, Yuan, Macri, \& Scolnic}]{Riess:2019cxk}
Riess, A.~G., Casertano, S., Yuan, W., Macri, L.~M., \& Scolnic, D. 2019, Astrophys. J., 876, 85, \dodoi{10.3847/1538-4357/ab1422}

\bibitem[{Riess {et~al.}(2022)Riess, Wenlong, Macri~Lucas, Dan, Dillon, Stefano, Jones~David, Yukei, Louise, Brink~Thomas, {et~al.}}]{riess2022comprehensive}
Riess, A.~G., Wenlong, Y., Macri~Lucas, M., {et~al.} 2022, Astrophys. J. Lett, 934, L7

\bibitem[{Rinaldi \& Del Pozzo(2021)}]{DelPozzononparam}
Rinaldi, S., \& Del Pozzo, W. 2021, Monthly Notices of the Royal Astronomical Society, 509, 5454, \dodoi{10.1093/mnras/stab3224}

\bibitem[{Sadiq {et~al.}(2022)Sadiq, Dent, \& Wysocki}]{adaptivekde}
Sadiq, J., Dent, T., \& Wysocki, D. 2022, Phys. Rev. D, 105, 123014, \dodoi{10.1103/PhysRevD.105.123014}

\bibitem[{Schutz(1986)}]{Schutz:1986gp}
Schutz, B.~F. 1986, Nature, 323, 310, \dodoi{10.1038/323310a0}

\bibitem[{Soares-Santos {et~al.}(2019)}]{DES:2019ccw}
Soares-Santos, M., {et~al.} 2019, Astrophys. J. Lett., 876, L7, \dodoi{10.3847/2041-8213/ab14f1}

\bibitem[{Suwa {et~al.}(2018)Suwa, Yoshida, Shibata, Umeda, \& Takahashi}]{Suwa:2018uni}
Suwa, Y., Yoshida, T., Shibata, M., Umeda, H., \& Takahashi, K. 2018, Mon. Not. Roy. Astron. Soc., 481, 3305, \dodoi{10.1093/mnras/sty2460}

\bibitem[{Talbot \& Thrane(2018)}]{Talbot:2018cva}
Talbot, C., \& Thrane, E. 2018, Astrophys. J., 856, 173, \dodoi{10.3847/1538-4357/aab34c}

\bibitem[{{The LIGO Scientific Collaboration} {et~al.}(2021){The LIGO Scientific Collaboration}, {The Virgo Collaboration}, \& {The KAGRA Collaboration}}]{gwtc-3pe}
{The LIGO Scientific Collaboration}, {The Virgo Collaboration}, \& {The KAGRA Collaboration}. 2021, {GWTC-3: Compact Binary Coalescences Observed by LIGO and Virgo During the Second Part of the Third Observing Run — Parameter estimation data release},  Zenodo, \dodoi{10.5281/zenodo.5546663}

\bibitem[{{The LIGO Scientific Collaboration} {et~al.}(2023){The LIGO Scientific Collaboration}, {Virgo Collaboration}, \& {KAGRA Collaboration}}]{gwtc-3inj}
{The LIGO Scientific Collaboration}, {Virgo Collaboration}, \& {KAGRA Collaboration}. 2023, {GWTC-3: Compact Binary Coalescences Observed by LIGO and Virgo During the Second Part of the Third Observing Run — O1+O2+O3 Search Sensitivity Estimates},  Zenodo, \dodoi{10.5281/zenodo.7890398}

\bibitem[{Tiwari(2021)}]{mixture1}
Tiwari, V. 2021, Classical and Quantum Gravity, 38, 155007, \dodoi{10.1088/1361-6382/ac0b54}

\bibitem[{Tiwari(2022)}]{structure2}
---. 2022, The Astrophysical Journal, 928, 155, \dodoi{10.3847/1538-4357/ac589a}

\bibitem[{Tiwari \& Fairhurst(2021)}]{structure1}
Tiwari, V., \& Fairhurst, S. 2021, The Astrophysical Journal Letters, 913, L19, \dodoi{10.3847/2041-8213/abfbe7}

\bibitem[{Vitale {et~al.}(2020)Vitale, Gerosa, Farr, \& Taylor}]{Vitale:2020aaz}
Vitale, S., Gerosa, D., Farr, W.~M., \& Taylor, S.~R. 2020, \dodoi{10.1007/978-981-15-4702-7_45-1}

\bibitem[{Woosley(2017)}]{Woosley:2016hmi}
Woosley, S.~E. 2017, Astrophys. J., 836, 244, \dodoi{10.3847/1538-4357/836/2/244}

\bibitem[{{Woosley}(2019)}]{2019ApJ...878...49W}
{Woosley}, S.~E. 2019, \apj, 878, 49, \dodoi{10.3847/1538-4357/ab1b41}

\bibitem[{Woosley {et~al.}(2002)Woosley, Heger, \& Weaver}]{Woosley:2002zz}
Woosley, S.~E., Heger, A., \& Weaver, T.~A. 2002, Rev. Mod. Phys., 74, 1015, \dodoi{10.1103/RevModPhys.74.1015}

\bibitem[{Ye \& Fishbach(2022)}]{Ye:2022qoe}
Ye, C., \& Fishbach, M. 2022, Astrophys. J., 937, 73, \dodoi{10.3847/1538-4357/ac7f99}

\end{thebibliography}
\bibliographystyle{aasjournal}

\end{document}